\documentclass[amsmath,amssymb, 12pt]{revtex4}
\usepackage{graphicx}
\usepackage{dcolumn}
\usepackage{bm}
\usepackage{graphicx}
\usepackage{epsfig}

\def\bea{\begin{eqnarray}}
\def\eea{\end{eqnarray}}

\begin{document}
\title{Holographic dark energy in the DGP model}
\author{Norman Cruz}
\altaffiliation{norman.cruz@usach.cl} \affiliation{Departamento de
F\'\i sica, Facultad de Ciencia, Universidad de Santiago, Casilla
307, Santiago, Chile.}
\author{Samuel Lepe}
\altaffiliation{slepe@ucv.cl} \affiliation{Instituto de F\'\i
sica, Facultad de Ciencias, Pontificia Universidad Cat\'olica de
Valpara\'\i so, Avenida Brasil 2950, Valpara\'\i so, Chile.}
\author{Francisco Pe\~na}
\altaffiliation{fcampos@ufro.cl} \affiliation{Departamento de
Ciencias F\'\i sicas, Facultad de Ingenier\'\i a, Ciencias y
Administraci\'on, Universidad de La Frontera, Avda. Francisco
Salazar 01145, Casilla 54-D Temuco, Chile.\\}
\date{\today}
\begin{abstract}
The braneworld model proposed by Dvali, Gabadadze and Porrati
leads to an accelerated universe without cosmological constant or
other form of dark energy. Nevertheless, we have investigated the
consequences of this model when an holographic dark energy is
included, taken the Hubble scale as IR cutoff. We have found that
the holographic dark energy leads to an accelerated universe flat
(de Sitter like expansion) for the two branch: $\epsilon =\pm 1$
of the DGP model. Nevertheless, in universes with no null
curvature the dark energy presents an EoS corresponding to a
phantom fluid during the present era and evolving to a de Sitter
like phase for future cosmic time. In the special case in which
the holographic parameter $c$ is equal to one we have found a
sudden singularity in closed universes. In this case the expansion
is decelerating.

\end{abstract}
\maketitle


\section{ Introduction}

The acceleration in the expansion of the universe during recent
cosmological times, first indicated by Supernovae
observations~\cite{Perlmutter} and also supported by the
astrophysical data obtained from WMAP, indicates, in the framework
of general relativity, the existence of an exotic fluid with
negative pressure, a form of matter which received the rather
confusing name of dark energy.  Other non conventional approaches
have advocated extra dimensions inspired in string and superstring
theories. One of this models that have leads to an accelerated
universe without cosmological constant or other form of dark
energy is the braneworld model proposed by Dvali, Gabadadze, and
Porrati (DGP)~\cite{DGP}, ~\cite{Deffayet},~\cite{Deffayet1} (for
reviews, see ~\cite{Koyama}). In a cosmological scenario, this
approach leads to a late-time acceleration as a result of the
gravitational leakage from a 3-dimensional surface (3-brane) to a
5-th extra dimension on Hubble distances.

More specifically, this model leads to an accelerated phase at
late times but with an effective dark energy component with
$w>-1$, and dependent of the redshift.  Since observations do not
exclude the possibility of crossing the phantom divide with an
effective energy density with an EoS with $w<-1$, recently Hirano
and Komiya~\cite{Hirano} have extended the modified Friedmann
equation proposed by Dvali and Turner~\cite{Dvali} , in order to
realize this crossing.

It is a well known fact that the DGP model has two branches of
solutions: the self-accelerating branch and the normal one. The
self accelerating branch leads to an accelerating universe without
invoking any exotic fluid, but present problems like
ghost~\cite{Koyama1}. Nevertheless, the normal branch requires a
dark energy component to accommodate the current
observations~\cite{Lue},~\cite{Lazkoz}. Extend models of gravity
on the brane with f(R) terms have been investigated to obtain self
acceleration in the normal branch~\cite{Mariam}. Solutions for a
DGP brane-world cosmology with a k-essence field were found
in~\cite{MariamChimento} showing big rip scenarios and
asymtotically de Sitter phase in the future.

The aim of the present work was to explore, in the framework of
the holographic dark energy models~\cite{Cohen}, ~\cite{Hsu},
~\cite{Li}, based on the holographic principle ~\cite{Gonzalez},
which is believed to be a fundamental principle for the quantum
theory of gravity, a DGP cosmology. We investigate the behavior of
a late time universe, considering dark energy as unique dominant
fluid.

Based on the validity of the effective quantum field theory, Cohen
et al~\cite{Cohen} suggested that the total energy in a region of
size $L$ should not exceed the mass of a black hole of the same
size, which means $ \rho_{\Lambda}\leq L^{-2}M_{p}^{2}$. The
largest $L$ is chosen by saturating the this bound so that we
obtain the holographic dark energy (HDE) density
\begin{eqnarray}
\rho_{\Lambda}=3c^{2}M_{p}^{2}L^{-2},\label{holobound}
\end{eqnarray}
where c is a free dimensionless $\mathcal{O}$(1) parameter that
can be determined by observations. Taking $L$ as the Hubble radius
$H = H_{0}^{-1}$ this $\rho_{\Lambda}$ is comparable to the
observed dark energy density, but gives wrong EoS for the dark
energy~\cite{Hsu}.

For higher dimensional space-times, the holographic principle in
cosmological scenarios has been formulated considering the maximal
uncompactified space of the model, i.e. in the bulk, leading to a
crossing of phantom divide for the holographic dark energy, in 5D
two-brane models~\cite{Saridakis}. Recently, a modified
holographic dark energy model has been formulated using the mass
of black holes in higher dimensions and the Hubble scale as IR
cutoff~\cite{Gong}. Using the future event horizon as IR cutoff,
it was found in that the EoS of the holographic dark energy can
cross phantom divide~\cite{Liu}.

We explore in this work the usual holographic bound in a DGP model
with curvature. We find that using the Hubble scale as IR cutoff,
the universe with the observable values corresponding to
curvature, the EoS of the dark energy can cross the phantom divide
and end in a de Sitter like expansion.

In the next section we present the DGP model for a universe with
curvature but without matter. In section III, the EoS of the
holographic dark energy with a Hubble cutoff is evaluated for
universes with no null curvature.  In section IV we discuss our
results.


\section{ DGP Model}
For an homogeneous and isotropic universe described by the
Friedmann-Robertson-Walker the field equation is given
by~\cite{Deffayet}, \cite{Deffayet1} (with 8$\pi G=1$)
\begin{eqnarray}\label{eq1-2}
3\left( H^{2}-\frac{\epsilon
}{r_{c}}\sqrt{H^{2}+\frac{k}{a^{2}}}\right) =\rho
-\frac{3k}{a^{2}},
\end{eqnarray}
where $H =\frac{\dot{a}}{a}$ is the Hubble parameter, $\rho$ is
the total cosmic fluid energy density on the brane, and $r_{c}$,
given by
\begin{eqnarray}\label{eq2}
r_{c}=\frac{1}{2}\frac{M_{\left( 4\right) }^{2}}{M_{\left(
5\right) }^{3}},
\end{eqnarray}
is a scale which sets a length beyond which gravity starts to leak
out into the bulk. The parameter $\epsilon =\pm 1$ represents the
two branches of the DGP model. It is well known that the solution
with $\epsilon = +1$ represent the self-accelerating branch, since
even without dark energy the expansion of the universe
accelerates, and for late times the Hubble parameter approaches a
constant, $H = 1/r_{c}$. In the previous investigation, $\epsilon
= -1$ has been named as the normal branch, where acceleration only
appears if a dark energy component is included.

Before to discuss the holographic behavior of dark energy in a DGP
model where no null curvature is taken account, we present briefly
the solutions without dark energy ( $\rho =0$). For this case
Eq.(\ref{eq1-2}) becomes
\begin{equation}
H^{2}+\frac{k}{a^{2}}=\frac{\epsilon
}{r_{c}}\sqrt{H^{2}+\frac{k}{a^{2}}},
\end{equation}
so only the cases  $k=\pm 1$ and $\epsilon =1$ admit physically
reasonable solutions. The equation to solve is then
\begin{equation}
H^{2}+\frac{k}{a^{2}}=\frac{1}{r_{c}^{2}}.
\end{equation}
which gives us the following solutions

\begin{equation}
a\left( t\right) =\left\{
\begin{array}{cc}
r_{c}\sinh \left( \frac{t}{r_{c}}+const.\right) , & k=-1, \\
r_{c}\cosh \left( \frac{t}{r_{c}}+const.\right) , & k=1.
\end{array}%
\right.
\end{equation}
The case $k=0$ correspond a de Sitter universe with $H=1/r_{c}$.

Let us consider a general  behaviors of the model for a flat
universe when one fluid is included, in terms of the size of the
quantity $r_{c}H$. Rewriting Eq.~(\ref{eq1-2}) using $r_{c}H$
yields
\begin{eqnarray}\label{eq4}
3H^{2}\left[ 1-\epsilon \left( r_{c}H\right) ^{-1}\right] &=&\rho.
\end{eqnarray}
If cosmic fluid satisfy a barotropic equation of state,
$p=\omega\rho$, the conservation equation is given by
\begin{eqnarray}\label{eq3}
\dot{\rho}+3H\left( 1+\omega \right) \rho &=&0.
\end{eqnarray}
From Eqs.~(\ref{eq4}) and (\ref{eq3}), we obtain an expression for
$\omega$ in terms of the Hubble parameter, which is given by
\begin{eqnarray}\label{eq5}
1+\omega =-\frac{1}{3}\left( \frac{2-\epsilon \left( r_{c}H\right)
^{-1}}{ 1-\epsilon \left( r_{c}H\right) ^{-1}}\right)
\frac{\dot{H}}{H^{2}}.
\end{eqnarray}
According to Eq.~(\ref{eq5}), $1+\omega <0$ implies that $
\dot{H}>0$, since $\left( 2-\epsilon \left( r_{c}H\right)
^{-1}\right) /\left( 1-\epsilon \left( r_{c}H\right) ^{-1}\right)
>0$, for both cases $\left( r_{c}H\right) ^{-1}\lessgtr 1$.

Using expressions ~(\ref{eq4}) and (\ref{eq5}), we find special
limits in terms of the size of $\left( r_{c}H\right) ^{-1}$. Let
see the three following cases: a) If $ \left( r_{c}H\right)
^{-1}<<1$, then $3H^{2}\rightarrow \rho$ and $1+\omega \rightarrow
-\frac{2}{3}\frac{\dot{H}}{H^{2}}$, which correspond to the
standard cosmology, in a four dimensional spacetime. b) If $
\left( r_{c}H\right) ^{-1}>>1$, then $3H\rightarrow r_{c}\rho$,
where we have assumed $\epsilon=-1$, so for a positive energy
density, the Hubble parameter is also positive. Independently of
the value of  $\epsilon$, we obtain for the equation of state,
$1+\omega \rightarrow -\frac{1}{3}\frac{\dot{H}}{H^{2}}$.


\section{ The holographic dark energy}

\subsection{ Hubble cut-off}

In what follows we shall consider an holographic dark energy,
which obeys the relation
\begin{eqnarray}\label{eq18}
\rho =3c^{2}H^{2}.
\end{eqnarray}

\textbf{Case $k=0$}. Using Eq.~(\ref{eq18}) in Eq.~(\ref{eq4}), a
direct solution for the Hubble parameter is obtained
\begin{eqnarray}\label{eq19}
H=\frac{\epsilon}{\left( 1-c^{2}\right) r_{c}},
\end{eqnarray}
along with an equation of state $\omega =-1$. In this scenario the
universe accelerates with a constant Hubble parameter, which
differs only by a factor $\epsilon /(1-c^{2})$ compared with the
self accelerating case, which appears in the branch $\epsilon =1$,
without dark energy. Note that the value of the holographic
parameter $c$ allows the following two cases, both with
acceleration: i) $ \epsilon =1$ and $c^{2}<1$, ii) $\epsilon =-1$
and $c^{2}>1$. The cases with $c\lessgtr 1$ were well discussed
from theoretical point of view in~\cite{Huang}, ~\cite{Horvat} and
from the observational point of view in~\cite{Huang1}.  Then, in
principle, both branch can behaves as a de Sitter universe.

In order to have appreciable modifications from the standard
cosmology at late times of cosmic evolution, it has been usually
assumed that $r_{c}\sim H_{0}^{-1}$ in the DGP model. So in the
holographic approach with the Hubble cutoff we obtain both
possible cases implies: i) $ \epsilon =1$ and $c^{2}\sim 0$, ii)
$\epsilon =-1$ and $c^{2}\sim 2$. From Eq.(\ref{eq18}), the first
case indicates that the holographic dark energy density is
approximately zero.  In other words, holography leads to obtain
the previous case of de Sitter universe without matter in the flat
case.  In the second one,  holography implies a de Sitter universe
with a high density of dark energy. In this sense,  the inclusion
of holography in the normal branch $\epsilon = -1$ leads to an
scenario with accelerated expansion but with mayor stuff of dark
energy.

\textbf{Cases with no null curvature}. For universes with non null
curvature, Eq.(\ref{eq1-2}) can be solved for the Hubble parameter
in terms of the variable $x=r_{c}/a$.

\begin{equation}\label{eq20}
\left[ \left( 1-c^{2}\right) r_{c}H_{\pm }\left( x\right) \right]
^{2}=\frac{ 1}{2}\left[ 1-\delta (x) \pm \sqrt{\Delta (x)}\right],
\end{equation}
where $\delta (x)= 2 k (1-c^{2})x^{2}$ and $\Delta (x)= 1-4kc^{2}
(1-c^{2})x^{2}$. In order to recover the flat case,  the physical
solution will be $H_{+}(x)$. Imposing the constraint $\Delta
(x)>0$ we obtain that $0\leq x^{2}< 1/4c^{2} (1-c^{2})$, for $k=1$
and $c^{2}<1$, and $0\leq x^{2}< 1/4c^{2} (c^{2}-1)$, for $k=-1$
and $c^{2}>1$. Note that these conditions implies $\delta (x)
>0$ and consequently $1-\delta (x) + \sqrt{\Delta (x)}>0$.

The constraints upon $x$ implies the that the scale factor $a$ has
the following lower bounds $a> 2cr_{c}\sqrt{1-c^{2}}$, or,  $a>
2cr_{c}\sqrt{c^{2}-1}$. In terms of the redshift these bounds are
$1+z < a_{0}/2cr_{c}\sqrt{1-c^{2}}$, or, $1+z <
a_{0}/2cr_{c}\sqrt{c^{2}-1}$

From Eq.(\ref{eq3}), and since $1+\omega \left( x\right) =\left(
2/3\right) \left( x/H\right) dH/dx$, and using Eq.(\ref{eq20}), we
obtain the following expression for the equation of state

\begin{equation}\label{omegaz}
1+\omega \left( x\right) =-\frac{4}{3}k(1-c^{2})\left( \frac{1+
c^{2}/2\sqrt{\Delta (x)}}{1-\delta (x) + \sqrt{\Delta
(x)}}\right)x^{2}
\end{equation}

Since the factor $k(1-c^{2})>0$, due to the conditions mentioned
above, Eq.(\ref{omegaz}) tell us that we have phantom evolution
for $ x>0$ (or $a>0$). Nevertheless,  a de Sitter phase is
achieved in the final evolution when $x=0$ (or $a=\infty$).

In order to compare our model with some observed values for the
present EoS of the dark energy, we will see the behavior of EoS,
given by Eq.(\ref{omegaz}), when $r_{c}\sim H_{0}^{-1}$.
Parameterizing the above condition, we can write $r_{c} H_{0}=
\alpha$, where $\alpha \sim 1$.  Redefining the variable $kx^{2}$
as $\eta (z)$, yields
\begin{equation}\label{eq23}
\eta (z)=\alpha^{2} \Omega_{k}(0)(1+z)^{2},
\end{equation}
where $\Omega_{k}(0)$ is the density parameter of the curvature at
the present time. Then in terms of the redshift Eq.(\ref{omegaz})
becomes
\begin{equation}\label{eq24}
1+\omega \left( z\right) =-\frac{4}{3}(1-c^{2})\left( \frac {\left
(1+ c^{2}/2\sqrt{1-4c^{2}(1-c^{2})\eta (z)}\right
)}{1-2c^{2}(1-c^{2})\eta (z)
 +
\sqrt{1-4c^{2}(1-c^{2})\eta (z)}} \right )
\end{equation}

The value of the holographic parameter $c^{2}$ has been fitted
with the current data for a variety of models and IR cutoff
(See~\cite{c}). Since in our case we are interested in the very
late time behavior we will use values for $c^{2}$ imposed above
for a closed an open universes. Using the last expression we have
plotted $\omega\left( z\right)$ in Fig. 1, for three values of the
parameter $\alpha$. We have considered the case of a closed
universe and take the value $\Omega_{k}(0)=0.0049$, which is
within the range given by WMAP observations (See~\cite{Hinshaw}).
The EoS of the holographic dark energy behaves like a phantom
fluid during the present time of the universe, but practically
indistinguishable of a cosmological constant. Fig. 2 shows a
similar behavior when $\alpha$ is fixed and the holographic
parameter is varied. Note, nevertheless, that for an open universe
the EoS correspond to a phantom fluid, with values within the
observational data, in the both cases with $c^{2}$ fixed (Fig. 3)
and with $\alpha$ fixed (Fig.4 ).

\begin{figure}[!h]
 \begin{center}
  \includegraphics[width=105mm]{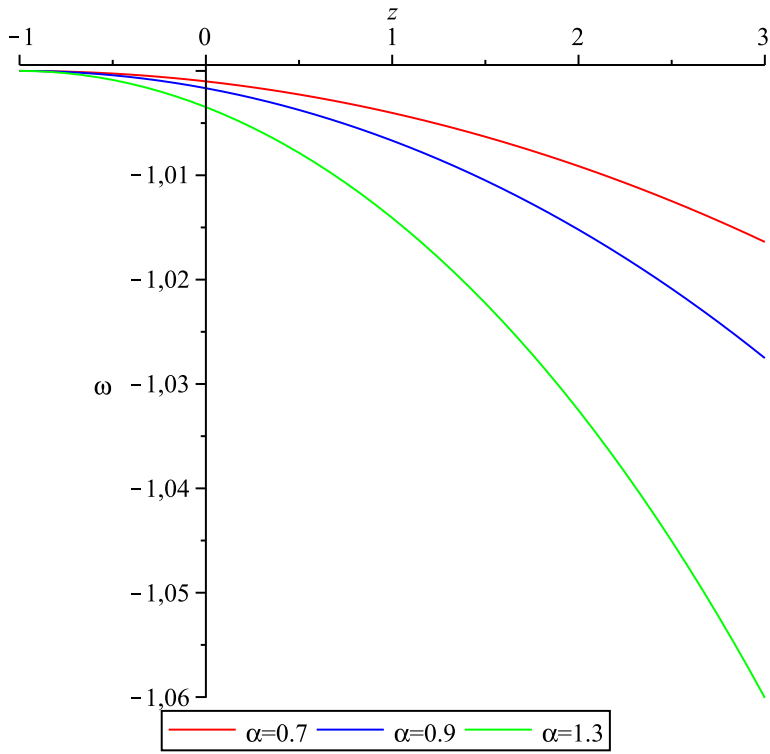}
 \end{center}
 \caption{EoS as a function of the redshift, for a closed universe, for three values of $\alpha$ and $c^{2}=0.25$.}
 \label{fig:fig.1}
\end{figure}



\begin{figure}[!h]
 \begin{center}
  \includegraphics[width=105mm]{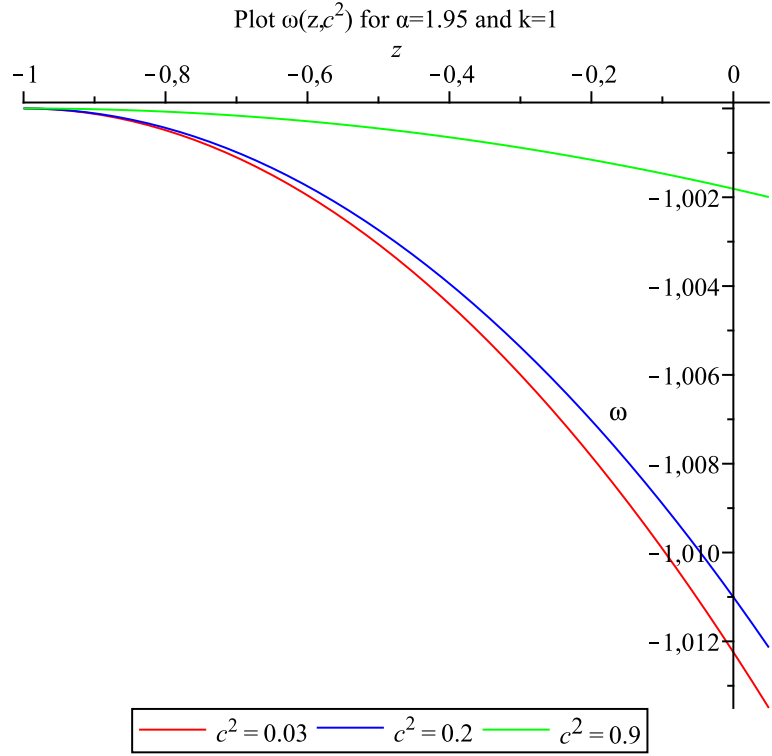}
 \end{center}
 \caption{EoS as a function of the redshift, for a closed universe, for three values of $c^{2}$ $\alpha=1.95$. }
 \label{fig:fig.1}
\end{figure}

\begin{figure}[!h]
 \begin{center}
  \includegraphics[width=105mm]{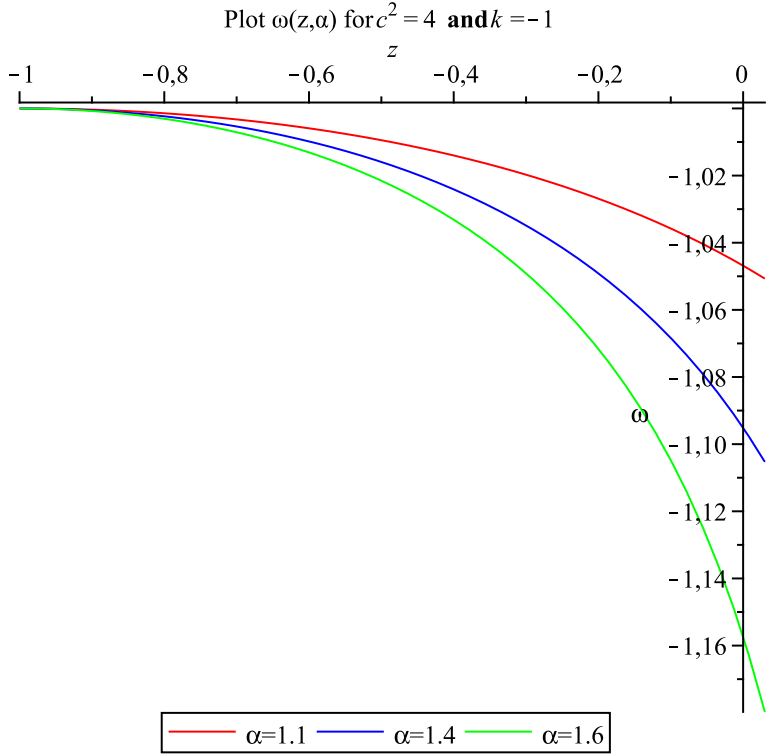}
 \end{center}
 \caption{EoS as a function of the redshift for an open universe, for three values of $\alpha$ and $c=4.0$. }
 \label{fig:fig.1}
\end{figure}

\begin{figure}[!h]
 \begin{center}
  \includegraphics[width=105mm]{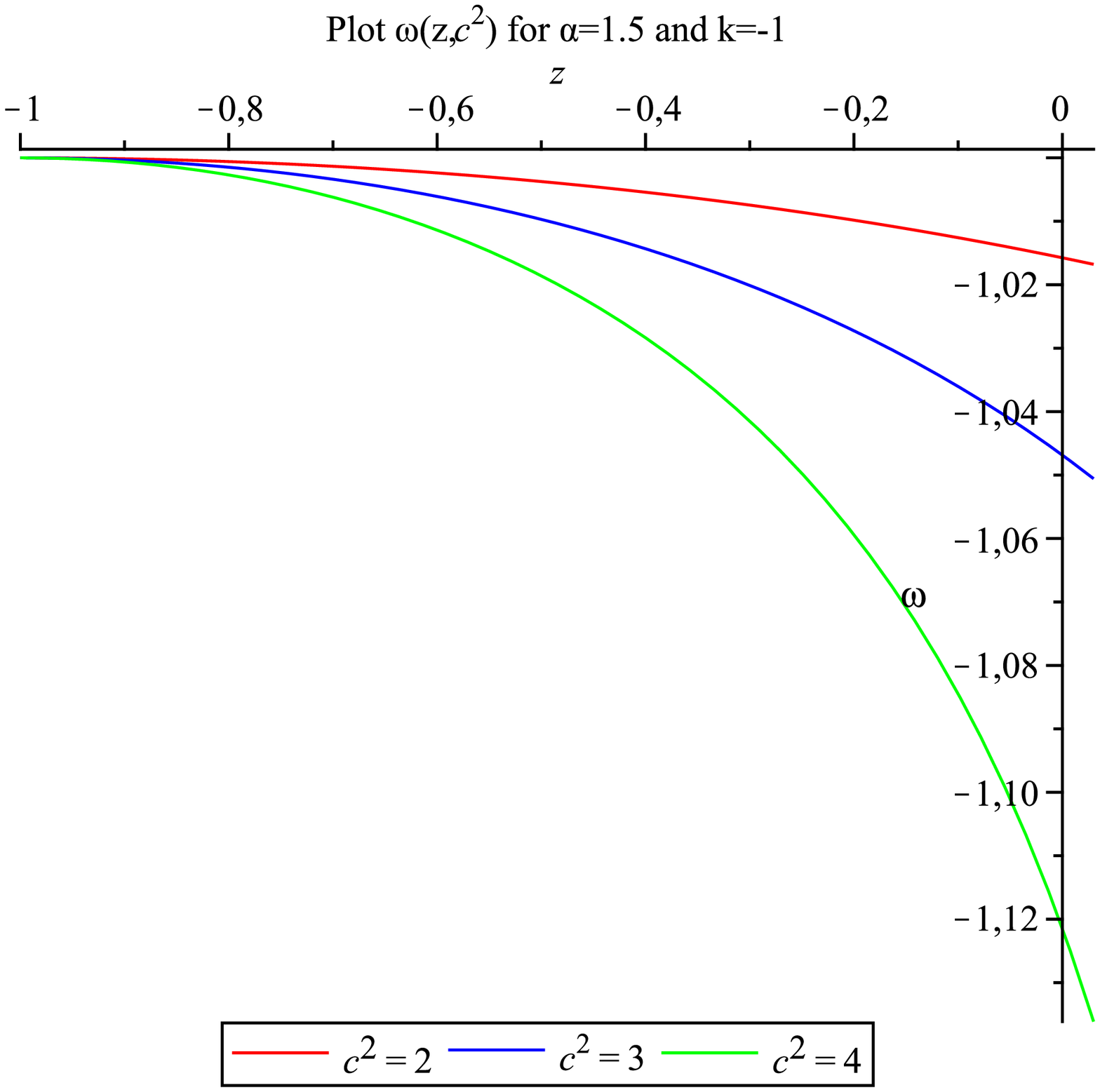}
 \end{center}
 \caption{EoS as a function of the redshift for an open universe, for three values of $c^{2}$ and fixed $\alpha$.}
 \label{fig:fig.1}
\end{figure}

\textbf{The special case} $\mathbf{c=1}$.   In this case, the
equation for $H$ becomes
\begin{equation}\label{eq20}
\frac{\epsilon }{r_{c}}\sqrt{H^{2}+\frac{k}{a^{2}}}=
\frac{k}{a^{2}},
\end{equation}
which has the following solution for the scale factor if the
initial condition is $a (t=t_{0})=a_{0}$
\begin{equation}\label{eq21}
ka^{2}(t)=r_{c}^{2}-(t_{s}-t)^{2},
\end{equation}
where we have chosen $\epsilon =1$. For a closed universe a sudden
singularity occurs at the time $t_{s}$, which is given by
\begin{equation}\label{eq22}
t_{s}=t_{0} + \sqrt{r_{c}^{2}-ka_{0}^{2}}.
\end{equation}
Using Eq.(\ref{eq21}) we can evaluate the EoS from the expression
$1+\omega =-2\dot{H}/H^{2}$
\begin{eqnarray}\label{eq23}
1+\omega =\frac{2}{3}\left( 1+
\frac{r_{c}^{2}}{(t_{s}-t)^{2}}\right).
\end{eqnarray}
The above equation indicates that the holographic dark energy has
an EoS with $\omega > -1/3$, indicating a non accelerated
universe, and a sudden singularity at $t=t_{s}$, since $1+\omega =
\infty$ and the scale factor is finite:
$a^{2}(t=t_{s})=r_{c}^{2}$. Additionally, the Hubble parameter,
given by
\begin{eqnarray}\label{eq24}
H^{2}(t) =\frac{1}{r_{c}^{2}-(t_{s}-t)^{2}}\left(
\frac{r_{c}^{2}}{r_{c}^{2}-(t_{s}-t)^{2}} -1\right),
\end{eqnarray}
is zero at $t=t_{s}$ a long with the dark energy density.  The
pressure is then finite at this time. Finally, the acceleration at
$t=t_{s}$ is $\ddot{a}/a =-1/r_{c}^{2}$.

\section{ Conclussions}

It is well known fact that the DGP models do not require a dark
energy to obtain an accelerated universe. Nevertheless,  we have
investigated the behavior of the cosmic evolution when an
holographic dark energy is included, using the Hubble scale as
cutoff.  We have considered the curvature in general,  obtaining
non trivial evolutions when the universes are closed or open.

Since the condition $r_{c}\sim H_{0}^{-1}$ is imposed in the DGP
framework to obtain only appreciable modifications to the standard
cosmology at late times, we have consider our results under this
condition.

For a flat universe, the holographic dark energy leads to an
accelerated universe (de Sitter like expansion) for the two
branch: $\epsilon =\pm 1$, under the following conditions for the
holographic parameter $c$: i) $ \epsilon =1$ and $c^{2}<1$, ii)
$\epsilon =-1$ and $c^{2}>1$. In this scenario the universe
accelerates with a constant Hubble parameter, which differs only
by a factor $\epsilon /(1-c^{2})$ compared with the self
accelerating case, which appears in the branch $\epsilon =1$,
without dark energy. Imposing the condition $r_{c}\sim H_{0}^{-1}$
we obtain that if we are in the branch $ \epsilon =1$ the
holographic dark energy density is approximately zero, which
implies that holography leads to obtain the previous case of de
Sitter universe without matter in the flat case. In the branch
$\epsilon = -1$, usually named as the normal branch, we obtain a
de Sitter universe with a high density of dark energy, which is
also similar to the results previous obtained where acceleration
only appears if a dark energy component is included.

In the cases with no null curvature we obtain the following
restrictions for the holographic parameter: i) if $k=1$, then
$c^{2}<1$, and ii) if $k=-1$, then $c^{2}>1$. The scale factor $a$
of these models has the following lower bounds  $1+z <
a_{0}/2cr_{c}\sqrt{1-c^{2}}$, and $1+z <
a_{0}/2cr_{c}\sqrt{c^{2}-1}$.  The main result of the holographic
DGP models with curvature is that the dark energy presents an EoS
corresponding to a phantom fluid during the present era and
evolving to a de Sitter like phase for future cosmic time. Our
results indicate that the holographic parameter $c$ has a little
influence in the present values of the phantom EoS of the dark
energy. Physically, it means that the increasing or decreasing of
the holographic dark energy is not a relevant factor in our model.
On the other hand, the condition $r_{c}\sim H_{0}^{-1}$, which was
parameterized throughout the parameter $\alpha$, measures the
modifications that induces the DGP approach on the standard field
equations at the present times. We have obtain that the phantom
behavior increases if $\alpha$ increases, or if $r_{c}\lesssim
H_{0}^{-1}$.

\section{acknowledgements}
NC and SL acknowledge the hospitality of the Physics Department of
Universidad de La Frontera where part of this work was done. SL
and FP acknowledge the hospitality of the Physics Department of
Universidad de Santiago de Chile. We acknowledge the support to
this research by CONICYT through grants Nos. 1110840 (NC) and
1110076 (SL). This work was also supported from DIUFRO DI10- 0009,
of Direcci\'on de Investigaci\'on y Desarrollo, Universidad de La
Frontera (FP) and DIR01.11,037.334/2011, VRIEA, Pontificia
Universidad Cat\'olica de Valpara\'\i so (SL).

\end{document}